\documentclass[showpacs,showkeys,twocolumn,preprintnumbers,nofootinbib,groupedaddress,superscriptaddress,amsmath,amssymb,showabstract]{revtex4-1}

\usepackage{graphicx}
\usepackage{dcolumn} 
\usepackage{bm}
\usepackage{amssymb}
\usepackage{amsmath}
\usepackage{epsfig}    
\usepackage{color}
\usepackage{slashed}
\usepackage{hyperref} 

\begin{document} 

\title{Gauged $U(1)_X$ breaking as origin of neutrino masses, dark matter and leptogenesis\\ at TeV scale}

\preprint{OU-HET-1172, KIAS-P23003}

\author{Toshinori Matsui}
\email{matsui@kias.re.kr}
\affiliation{School of Physics, KIAS, Seoul 02455, Korea}
\thanks{Address after April 2023, National Institute of Technology, Kure College.}

\author{Takaaki Nomura}
\email{nomura@scu.edu.cn}
\affiliation{College of Physics, Sichuan University, Chengdu 610065, China}

\author{Kei Yagyu}
\email{yagyu@het.phys.sci.osaka-u.ac.jp}
\affiliation{Department of Physics, Osaka University, Toyonaka, Osaka 560-0043, Japan}

\begin{abstract}

We propose a new mechanism which simultaneously explains 
tiny neutrino masses, stability of dark matter and baryon asymmetry of the Universe via leptogenesis due to the common origin: a spontaneous breaking of a $U(1)_X$ gauge symmetry at TeV scale. 
The $U(1)_X$ breaking provides small Majorana masses of vector-like leptons which generate small mass differences among them, and 
enhance their CP-violating decays via the resonant effect. 
Such CP-violation and lepton number violation turns out to be a sufficient amount of the observed baryon asymmetry through leptogenesis. 
The Majorana masses from the $U(1)_X$ breaking also induce radiative generation of masses for active neutrinos at one-loop level. 
Furthermore, a $Z_2$ symmetry appears as a remnant of the $U(1)_X$ breaking, which guarantees the stability of dark matter. 
We construct a simple renormalizable model to realize the above mechanism, and show a benchmark point which can explain observed neutrino oscillations, dark matter data and the baryon asymmetry at the same time.



\end{abstract}
\maketitle

\noindent
{\it Introduction ---}
Neutrino oscillations, existence of dark matter and baryon asymmetry of the Universe
have been well known and established phenomena which cannot be explained in the standard model (SM) of particle physics. 
Thus, there is no doubt about the necessity for new physics beyond the SM. 
So far, plethora of models have been proposed to explain these phenomena, some of which can simultaneously explain all of them. 

One of the simplest such new physics models is that with right-handed neutrinos.
Masses and mixings of active neutrinos can be explained by the type-I seesaw mechanism~\cite{Minkowski:1977sc,Gell-Mann:1979vob,Yanagida:1979as,Mohapatra:1979ia}.
Assuming one of the right-handed neutrinos to be odd under a $Z_2$ symmetry, it can be a candidate of dark matter. 
In addition, decays of the $Z_2$-even right-handed neutrinos can generate CP-violation and the lepton number, which can be converted into the baryon asymmetry of the Universe via the sphaleron process~\cite{Kuzmin:1985mm}, 
that is the leptogenesis scenario~\cite{Fukugita:1986hr}. 
In this scenario, right-handed neutrinos ``unify'' the explanation of three phenomena at the same time. 
Although this simple model works well, its experimental probe is generally quite challenging, 
because masses of the right-handed neutrinos typically have to be of order $10^{10}$ GeV or larger, see e.g.,~\cite{Strumia:2006qk}. 
Furthermore, the $ad~hoc$ $Z_2$ symmetry is not originated from dynamics. 

In this Letter, we propose a new mechanism which simultaneously explains 
tiny neutrino masses, stability of dark matter and the baryon asymmetry via leptogenesis, 
in which all of them is originated from a spontaneous breaking of a $U(1)_X$ gauge symmetry at TeV scale. 
In our scenario, vector-like leptons with non-zero $U(1)_X$ charges are introduced, which have Dirac masses at tree level. 
After the $U(1)_X$ breaking, Majorana masses for these vector-like leptons appear, by which 
small mass differences are generated in their mass eigenstates. 
Such a mass difference can enhance CP-violating (CPV) decays of the heavy neutral leptons due to the resonant effect~\cite{Pilaftsis:1997jf,Pilaftsis:2003gt}, and then 
sufficient amount of the baryon asymmetry is explained via the leptogenesis. 
The $U(1)_X$ breaking also provides a $Z_2$ symmetry as a remnant, by which stability of the lightest $Z_2$-odd particle is guaranteed, and it can be a candidate of dark matter. 
Furthermore, the Majorana masses for the vector-like leptons and the $Z_2$ symmetry 
realize the so-called scotogenic mechanism~\cite{Ma:2006km}, where tiny masses for active neutrinos are generated at one-loop level. 
Effectively, our scenario is similar to the scotogenic model with the low-scale leptogenesis~\cite{Hugle:2018qbw,Borah:2018rca,Mahanta:2019gfe,Sarma:2020msa,Kashiwase:2013uy,Chun:2020vxo}, but the CPV decay of heavy Majorana fermions is enhanced by not only the resonant effect but also sizable Yukawa couplings associated with a scalar field whose vacuum expectation value (VEV) breaks the $U(1)_X$ symmetry. 

In the following, we construct a simple model to realize the above-mentioned mechanism, and give successful benchmark points to explain current neutrino data and the observed baryon asymmetry of the Universe. 

\begin{table}[t]
\begin{center}
\begin{tabular}{|c||cccc|}\hline
Fields    &  $N^{a}$   & $\eta$       &  $\chi$   & $\varphi$    \\\hline\hline
$SU(2)_L$ &  ${\bm 1}$       &  ${\bm 2}$   & ${\bm 1}$ & ${\bm 1}$    \\\hline
$U(1)_Y$  &  $0$             &  $1/2$       & $0$       & 0  \\\hline
$U(1)_X$  &  $1$            &  $1$         & $1$      & $-2$  \\\hline
\end{tabular}
\caption{Charge assignments under the $SU(2)_L\times U(1)_Y \times U(1)_X$ gauge symmetry, where $N^a$ ($a=1,2$) are the vector-like leptons and all the others are scalar fields. }
\label{tab:particle}
\end{center}
\end{table}

{\it Model --} 
The content of new fields is shown in Table~\ref{tab:particle}, where all the SM fields have the same quantum number as those in the SM. \footnote{Setup of our model 
is similar to that given in Ref.~\cite{Chun:2020vxo}, in which an isospin triplet scalar field is introduced to close the $\eta$-loop in the one-loop diagram for neutrino masses. 
In our scenario, the $\mu_1$ and $\mu_2$ terms given in Eq.~(\ref{eq:potential}) play the similar role without introducing triplets. 
In addition, in~\cite{Chun:2020vxo}, the lepton number asymmetry is mainly produced via the resonant $Z'$ effect in two-body to two-body scatterings, while
such scatterings are negligibly small in our model, and the lepton number asymmetry is mainly produced via the decay of heavy Majorana fermions. }
The relevant new terms in the following discussion are given by 
\begin{align}
-&{\cal L}_{\rm rel}  
=  M_a \overline{N_L^a}N_R^a + y_\eta^{ia} \overline{ L_L^i} (i\sigma_2\eta^*) N_R^a + \text{h.c.}, \notag\\
&+ y_L^{ab} \overline{N_L^{ac}}N_L^{a}\varphi + y_R^{ab} \overline{N_R^{ac}}N_R^a \varphi + \text{h.c.}, \notag\\
&+ \sum_{\Phi=\eta,\chi,\varphi}m_\Phi^2|\Phi|^2  + \left(\mu_1 \eta^\dagger H \chi + \frac{\mu_2}{2} \varphi\chi\chi  + \text{h.c.}\right) , \label{eq:potential}
\end{align}
where $L_L^i$ ($i=1$--3) and $H$ are the $i$-th generation of the SM lepton doublet and the Higgs doublet, respectively. 
The superscript $c$ denotes the charge conjugation, and $\sigma_2$ is the second Pauli matrix. 
The Dirac masses $M_a$ ($a=1$,2) can be taken to be diagonal with real and positive values by the bi-unitarity transformation of $N^a$. 
In this basis, the Yukawa matrices $y_\eta$ and $y_{L,R}$ are generally complex, where the latter are symmetric due to the $SU(2)_L$ structure. 
The phases of $\mu_1$ and $\mu_2$ can be removed by rephasing $\chi$ and $\eta$ without loss of generality. 

It is important to mention here that a non-zero value of $\mu_1\mu_2$ explicitly breaks the global lepton number symmetry $U(1)_L$. 
In other words, if we take $\mu_1$ and/or $\mu_2$ to be zero, the theory recovers the $U(1)_L$ symmetry, in which 
the lepton number of $\varphi$ can be taken to be an arbitrary value by choosing those of $\eta$, $\chi$ and $N^a$ appropriately. 
This means that the $U(1)_L$ symmetry becomes $exact$ for the case with $\mu_1\mu_2 = 0$, and thus Majorana masses for 
the left-handed neutrinos vanish as we will discuss it soon below.

The $U(1)_X$ symmetry is broken down by the VEV $\langle\varphi\rangle = v_\varphi/\sqrt{2}$. 
Assuming the VEVs of $\eta$ and $\chi$ to be zero, a $Z_2$ symmetry remains as the remnant of the $U(1)_X$ symmetry, where 
fields with an odd number of the $U(1)_X$ charge, i.e., $N^a$, $\eta$ and $\chi$ are $Z_2$-odd while all the other fields are even. 
Then, the lightest $Z_2$-odd particle can be a candidate of dark matter. 

The neutral components of the $Z_2$-odd scalars, 
$\eta^0 = (\eta_H + i \eta_A)/\sqrt{2}$ and $\chi = (\chi_H + i \chi_A)/\sqrt{2}$, 
are mixed with each other due to the $\mu_1$ term. 
We define the mass eigenstates of these scalar fields as 
\begin{align}
\begin{pmatrix}
\eta_X^{}\\
\chi_X^{}
\end{pmatrix}=
\begin{pmatrix}
\cos\theta_X & -\sin\theta_X \\
\sin\theta_X & \cos\theta_X
\end{pmatrix}
\begin{pmatrix}
X_1\\
X_2
\end{pmatrix}, \label{eq:mass}
\end{align}
with $X = H,~A$. We note that in the limit 
$\mu_1 \to 0$, these mixing angles become zero and $m_{H_1} = m_{A_1}$, while 
in the limit $\mu_2 \to 0$, these mixing angles become identical and 
$m_{H_i} = m_{A_i}$ ($i = 1,2$).   

After the $U(1)_X$ breaking, $N^a$ obtain the Majorana masses and their mass term is expressed as
\begin{align}
{\cal L}_{\rm mass} = -\frac{1}{2}\overline{\tilde \Psi_R^{c}} \tilde{M}_\Psi \tilde \Psi_R + \text{h.c.}, 
\end{align}
where $\tilde \Psi_R \equiv (N^{1c}_L, N_R^1, N^{2c}_L,N_R^2)^T$, and $\tilde{M}_\Psi$ is the $4\times 4$ mass matrix given as
\begin{align}
 \tilde{M}_\Psi  = 
\begin{pmatrix} 
(\delta m_L^*)_{11} & M_1 & (\delta m_L^*)_{12} & 0 \\
 M_1 & (\delta m_R)_{11}  & 0 & (\delta m_R)_{12}  \\
 (\delta m_L^*)_{12} & 0 & (\delta m_L^*)_{22} &  M_2 \\
 0 & (\delta m_R)_{12} & M_2 & (\delta m_R)_{22} 
 \end{pmatrix}. \label{eq:matrix}
\end{align}
In the mass matrix, we introduce $\delta m_{L,R} \equiv \sqrt{2} v_\varphi y_{L,R}$. 
We can diagonalize the mass matrix by introducing the $4 \times 4$ unitary matrix $V$ as $M_\Psi \equiv V^T \tilde{M}_\Psi V = \text{diag}(m_{\psi_1},m_{\psi_2},m_{\psi_3},m_{\psi_4})$ with $m_{\psi_4}\geq m_{\psi_3}\geq m_{\psi_2}\geq m_{\psi_1}$.
The mass eigenstates are then given by 
\begin{equation}
\Psi \equiv (\psi_1, \psi_2, \psi_3, \psi_4)^T = V \tilde \Psi. 
\end{equation}

 \begin{figure}[!t]
 \begin{center}
  \includegraphics[width=50mm]{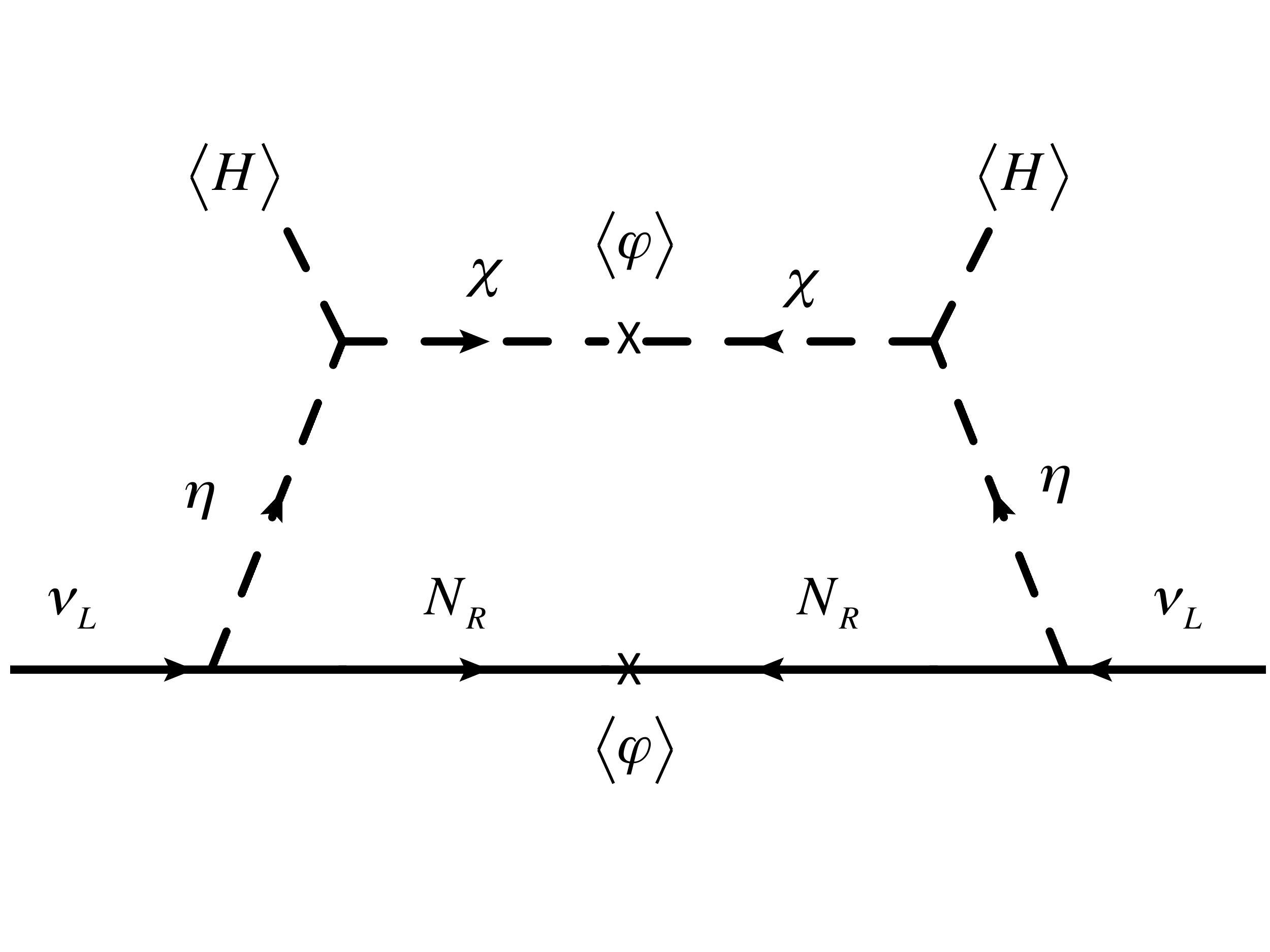}
    \caption{One-loop generation of neutrino masses. }
    \label{fig:diagram}
 \end{center}
 \end{figure}

\begin{figure}[!t]
 \begin{center}
  \includegraphics[width=50mm]{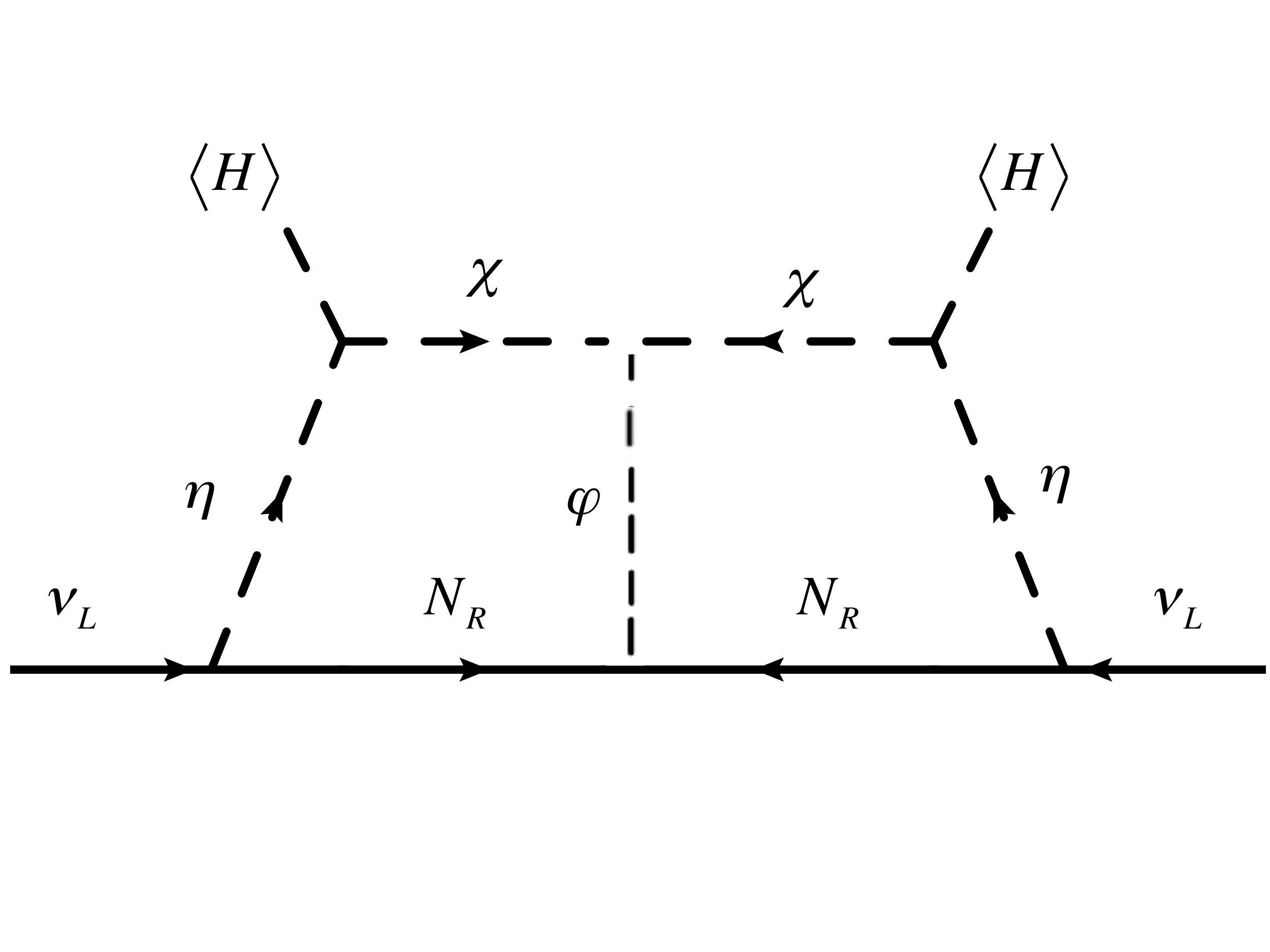}
    \caption{Example of higher loop contributions to neutrino masses which do not vanish in the limit $v_\varphi \to 0$. }
    \label{fig:nudiagram2}
 \end{center}
 \end{figure}

{\it Neutrino mass --} 
Majorana masses for active neutrinos are generated from the one-loop diagram shown in Fig.~\ref{fig:diagram}. 
The mass matrix is calculated as 
\begin{align}
& (m_\nu )_{ij} = \sum_{I=1}^4 
\frac{ Y^{*iI}_\eta  Y^{*jI}_\eta}{32 \pi^2} m_{\psi_I}  \sum_{X=H,A} p^X \notag\\
&\left( \frac{m_{X_1}^2 \cos^2\theta_X }{m^2_{X_1} - m_{\psi_I}^2} \ln \frac{m^2_{X_1}}{m^2_{\psi_I}}
+\frac{ m_{X_2}^2\sin^2\theta_X}{m^2_{X_2} - m_{\psi_I}^2} \ln \frac{m^2_{X_2}}{m^2_{\psi_I}}
\right), \label{eq:massmat}
\end{align}
where $p^H~(p^A) = 1~(-1)$, and $Y_\eta^{iI} = y_\eta^{i1}V^{2I} + y_\eta^{i2}V^{4I}$. 
We note that the above matrix vanishes for $\mu_1\mu_2 = 0$, because the $U(1)_L$ symmetry is recovered in this limit. 
This can explicitly be shown by using the properties mentioned just below Eq.~(\ref{eq:mass}). 
Therefore, both $\mu_1$ and $\mu_2$ should be non-zero in order to obtain finite neutrino masses. 
In addition, the mass matrix also vanishes in the limit of $v_\varphi \to 0$. 
Although this can be seen by looking at Fig.~\ref{fig:diagram}, but we can explicitly show it as follows. 
In this limit, the mass matrix for $\tilde{\Psi}_R$ given in Eq.~(\ref{eq:matrix}) becomes a block diagonal form, and 
we obtain $m_{\psi_1} = m_{\psi_2}$ and $m_{\psi_3} = m_{\psi_4}$. At the same time, the unitary matrix $V$ becomes a simple form of 
\begin{align}
V \to \frac{1}{\sqrt{2}} 
\begin{pmatrix}
-i & 1 & 0 & 0 \\
 i & 1 & 0 & 0 \\
 0 & 0 &-i & 1 \\
 0 & 0 & i & 1 
\end{pmatrix}. 
\end{align}
Using the above matrix and the mass degeneracy, we can show that the contributions from $\psi_1$ and $\psi_2$ ($\psi_3$ and $\psi_4$) are exactly cancelled. 
This, however, does not mean that active neutrinos become massless, because the $U(1)_L$ symmetry is explicitly broken at Lagrangian level as mentioned above. 
In fact, we can find higher loop contributions to the Majorana mass for active neutrinos, and one of such examples is shown in Fig.~\ref{fig:nudiagram2}. 
Throughout this Letter, we do not take into account such higher loop contributions, and suppose that the one-loop contribution given in Eq.~(\ref{eq:massmat}) is dominant. 
We also note that our mass matrix has rank 2, so that the lightest neutrino becomes massless. 

%


 \begin{figure}[!t]
 \begin{center}
  \includegraphics[width=80mm]{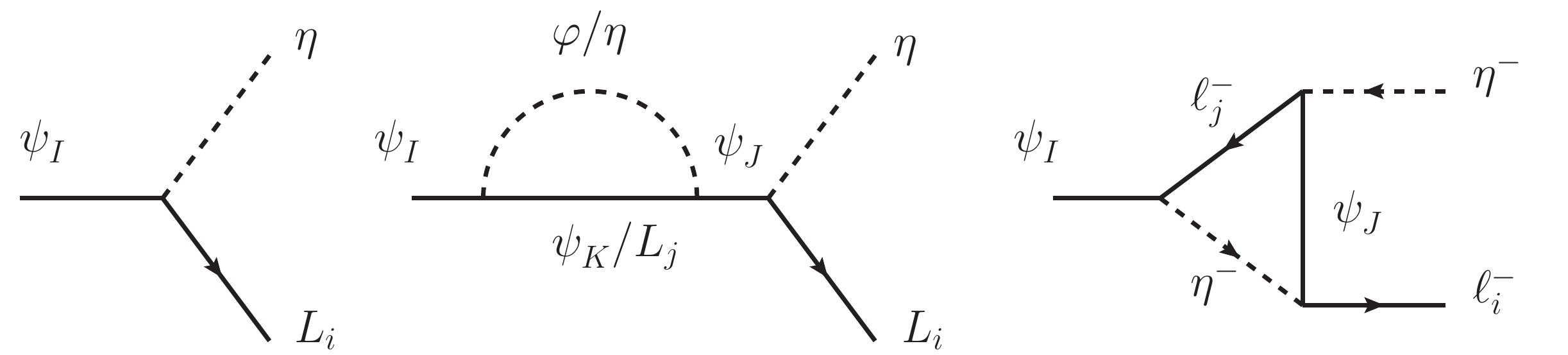}
    \caption{CP-violating decays of $\psi_I^{}$. }
    \label{fig:diagram2}
 \end{center}
 \end{figure}

{\it Leptogenesis --} In our scenario, the lepton number density $n_L$ or the $B-L$ number density $n_{B-L}$
can be generated through the decay of the Majorana fermions $\psi_I$ shown in Fig.~\ref{fig:diagram2},  
if we take non-zero CPV phases of the Yukawa couplings, and if the decay occurs in the out-of-thermal equilibrium.  
The produced lepton number is then converted into the baryon number density $n_B$ via the sphaleron process~\cite{Kuzmin:1985mm} according to the following equation~\cite{Khlebnikov:1988sr}
\begin{align}
n_B^{} &= \frac{8}{23}n_{B-L}^{}, 
\end{align}
which is derived by using the relation given by the chemical equilibrium for the sphaleron process, Yukawa interactions (except for that with the vector-like leptons) and conservation of the hypercharge. 
For the discussion of leptogenesis, we neglect the mixing effect shown in Eq.~(\ref{eq:mass}) for simplicity, which does not essentially change the conclusion. 

We first consider the out-of-equilibrium decay of $\psi_I$ whose amount can be described by introducing the following efficiency parameter $K_I$~\cite{Kolb:1990vq}: 
\begin{align}
K_I \equiv \frac{\langle \Gamma(\psi_I \to L\eta) \rangle_{T = m_{\psi_I}}}{2H(m_{\psi_I})} \sim \frac{|\bar{Y}_\eta|^2}{16\pi\sqrt{g_*}}\frac{M_{\rm pl}}{m_{\psi_I}}, \label{eq:k}
\end{align}
where $H(m_{\psi_I})$ is the Hubble parameter at the temperature $T$ to be $m_{\psi_I}$, 
$\bar{Y}_{\eta}$ is the averaged value of the Yukawa couplings $Y_\eta^{Ii}$, $M_{\rm pl}=1.22\times 10^{19}$ GeV is the Planck mass, and 
$g_* \simeq 124$ is the effective massless degrees of freedom assuming all the particles in our model being massless. 
In Eq.~(\ref{eq:k}), we introduced the thermally averaged decay rate defined as 
\begin{align}
&\langle \Gamma(\psi_I\to L\eta) \rangle \notag\\
&\equiv \sum_{\rm fin}[\Gamma(\psi_I \to L\eta ) + \Gamma(\psi_I \to L^c \eta^c )]\frac{{\cal K}_1(m_{\psi_I}/T)}{{\cal K}_2(m_{\psi_I}/T)}, 
\end{align}
where $\Gamma(\psi_I \to XY)$ denotes the decay rate of $\psi_I \to XY$ 
with $\sum_{\rm fin}$ denoting the summation for all the possible final states, i.e., isospin components and lepton flavors, and 
${\cal K}_{n}(z)$ are the modified Bessel functions of the $n$-th kind.
This $K_I$ parameter can be of order one, i.e., the decay rate is compatible with the expansion rate of the Universe and provide sizable amount of out-of thermal equilibrium, 
for $m_{\psi_I} = {\cal O}(10)$ TeV and $\bar{Y}_\eta = {\cal O}(10^{-6})$. 
However, to reproduce the active neutrino masses to be ${\cal O}(0.1)$ eV, 
$\bar{Y}_\eta$ has to be of order $10^{-4}$ or larger, so that typically we obtain $K_I > 10^4$, which corresponds to the so-called strong wash-out regime. 
The yield for the baryon number $Y_B \equiv n_B^{}/s$ with $s$ being the entropy density 
is roughly estimated as $Y_B \sim -8/23 \times 0.3\epsilon/[g_*K(\ln K)^{0.6}]$~\cite{Kolb:1990vq} with $\epsilon$ and $K$ to be max$(|\epsilon_I|)$ describing the amount of CP-violation
defined in Eq.~(\ref{eq:eps1}) and the corresponding $K_I$ value, respectively. \footnote{This expression gives a good approximation particularly for $K = K_I \gg K_{I'}$ with $m_{\psi_I} > m_{\psi_{I'}}$. 
}
Thus, we need a larger value of $\epsilon$ parameter, typically ${\cal O}(10^{-2})$, to compensate the suppression factor of $1/K$. 
For the actual calculation of $Y_B$, we numerically solve the Boltzmann equations, discussed below. 

Next, we discuss the CPV decay of $\psi_I$. 
As in the ordinal leptogenesis scenario, the CPV effect appears from the interference between 
the tree diagram and the one-loop diagrams shown in Fig.~\ref{fig:diagram2} at leading order. 
The amount of the CP-violation is expressed by introducing the following asymmetric parameter: 
\begin{align}
\epsilon_I &\equiv \frac{\sum_{\rm fin}[\Gamma(\psi_I \to L\eta) - \Gamma(\psi_I \to L^c \eta^c)]}{\sum_{\rm fin}[\Gamma(\psi_I \to L\eta ) + \Gamma(\psi_I \to L^c \eta^c )]}. \label{eq:eps1} 
\end{align}
The magnitude of these $\epsilon_I$ parameters is determined by two types of the Yukawa couplings, i.e., $y_\eta$ and $y_{L,R}$. 
As aforementioned, the magnitude of $y_\eta$ has to be of order $10^{-4}$ to reproduce the active neutrino masses and to avoid
a too strong wash-out of the generated lepton number, while $y_{L,R}^{}$ can be of order one. 
Thus, the contribution from the $\varphi$ loop shown in Fig.~\ref{fig:diagram2} is dominated with respect to the $\eta$ loop one, 
and hence we can safely ignore the vertex correction shown as the third diagram. 
The self-energy diagram (the second one in Fig.~\ref{fig:diagram2}) 
can also be enhanced by using the resonant effect of the intermediate $\psi_J$ if 
a small mass difference between $\psi_I^{}$ and $\psi_J$ is taken, because the amplitude is proportional to $(m_{\psi_I}^2 - m_{\psi_J}^2)^{-1}$. 
We note that in order to make the contribution from the self-energy diagram with $\varphi$-loop non-zero, the sum of the masses of $\psi_K^{}$ and $\varphi$ must be smaller than
that of $\psi_I^{}$, because a non-zero value of $\epsilon_I$ requires both ``weak phase'' coming from the imaginary part of 
the Yukawa coupling and the ``strong phase'' coming from loop functions. 
The latter becomes non-zero when the particles in the loop are on-shell. 

 \begin{figure}[!t]
 \begin{center}
  \includegraphics[width=60mm]{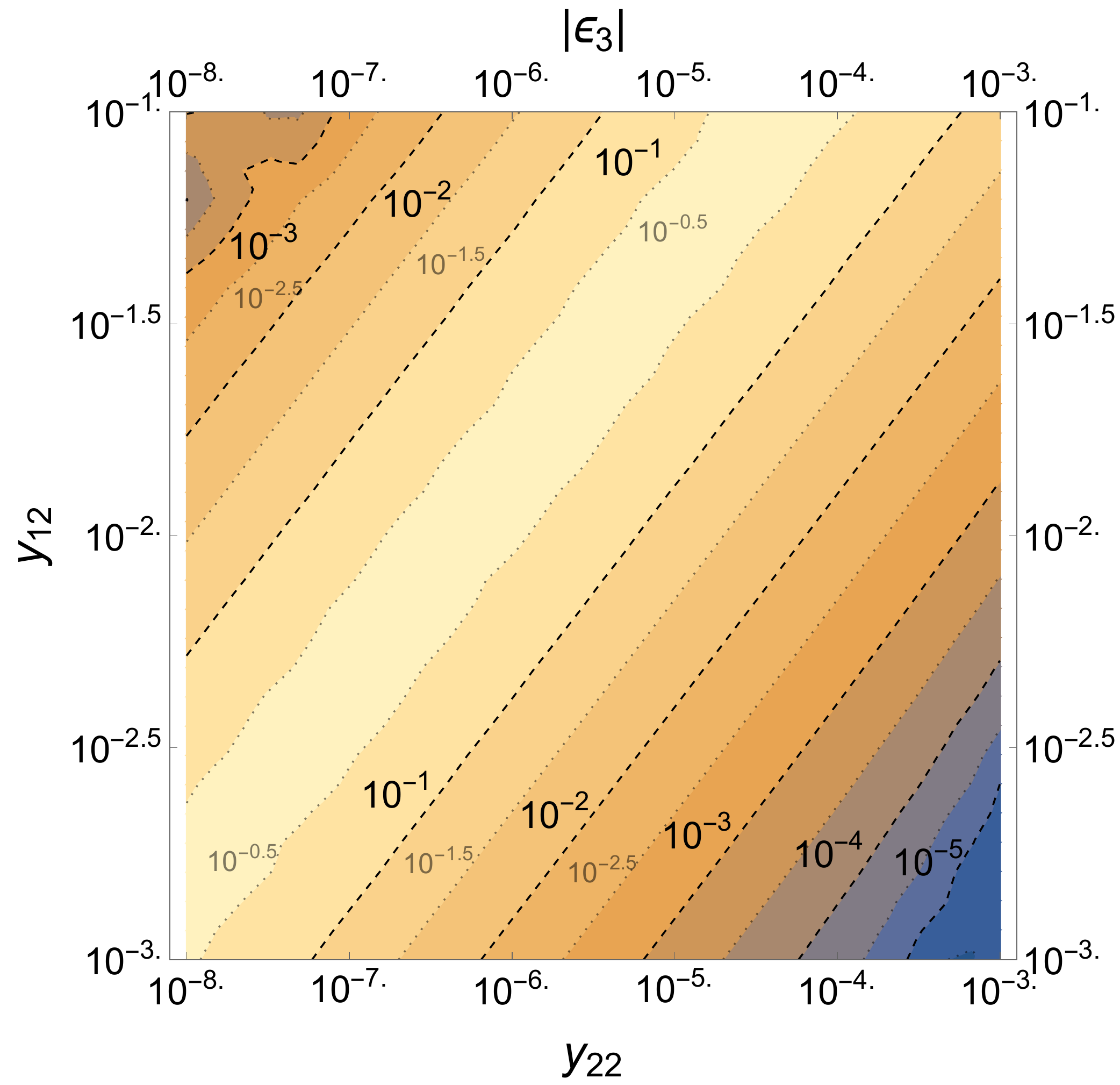}
    \caption{Contour plot for $|\epsilon_3|$ with $y_\eta = 10^{-4}$ and $r=1$.}
    \label{fig:num1}
 \end{center}
 \end{figure}

 \begin{figure}[!t]
 \begin{center}
  \includegraphics[width=60mm]{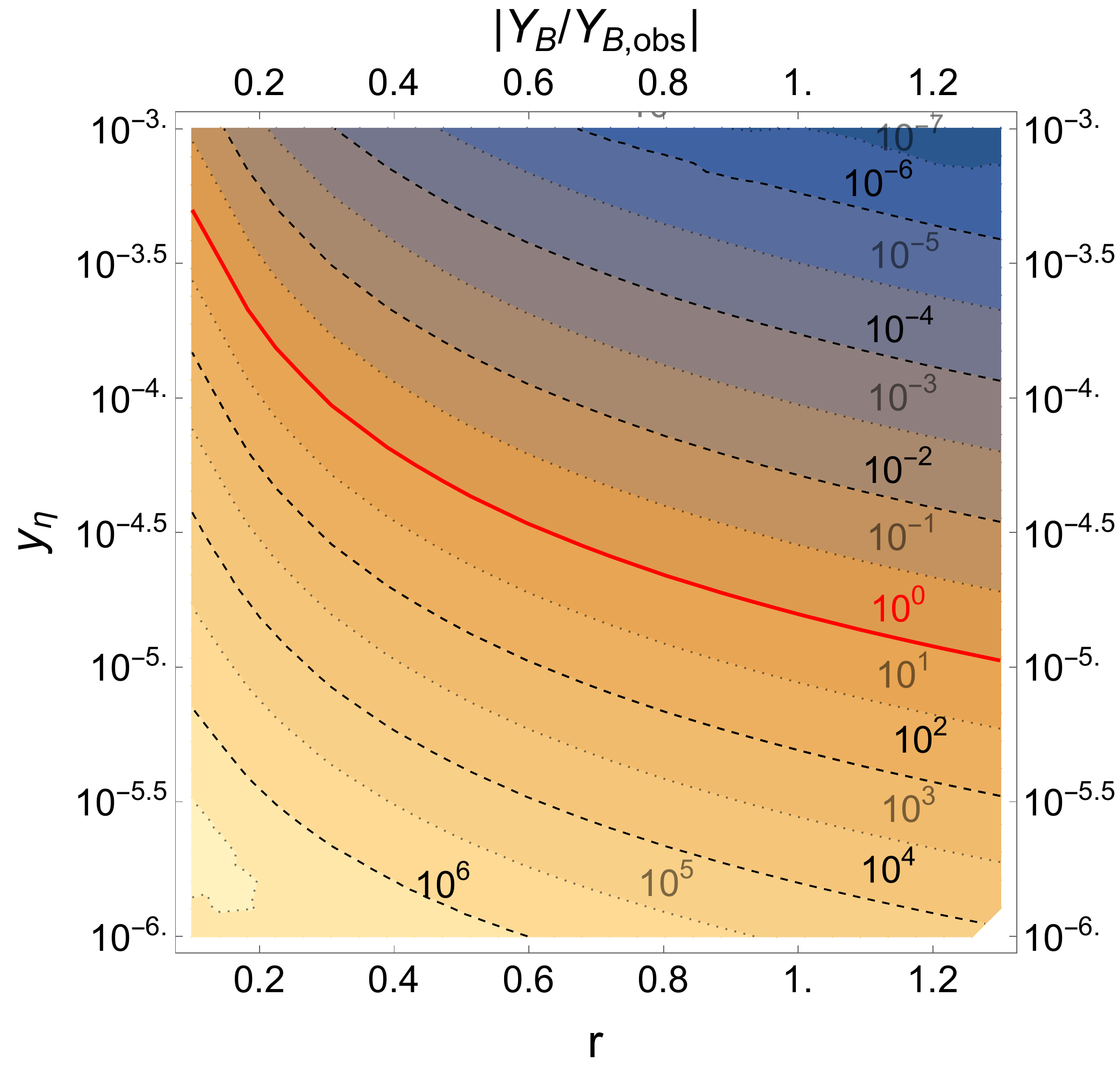}
    \caption{Contour plot for $|Y_B|/Y_B^{\rm obs}$  with $y_{12}=10^{-1.5}$ and $y_{22}=10^{-5}$. 
}
    \label{fig:num2}
 \end{center}
 \end{figure}

{\it Numerical results --} To evaluate $Y_B$, we numerically solve the following set of the Boltzmann equations:
\begin{align}
\frac{dY_I}{dz} &=   -\frac{z}{H(m_{\psi_1})}\Bigg[\langle \Gamma(\psi_I \to  L \eta)\rangle \left(Y_{I} - Y_{I}^{\rm eq} \right)    \notag\\
&  +\sum_{J \neq I}\langle \Gamma(\psi_I \to  \psi_J\varphi)\rangle \left(Y_I - \frac{Y_I^{\rm eq}}{Y_J^{\rm eq}}\,Y_J \right) \Bigg] , \\
\frac{dY_L}{dz} &=   \frac{z}{H(m_{\psi_1})}\sum_I\langle \Gamma(\psi_I \to  L\eta)\rangle \notag\\
& \times \left[(Y_I - Y_I^{\rm eq})\,\epsilon_I - \frac{Y_L}{2Y_{\rm rel}^{\rm eq}}\,Y_{I}^{\rm eq} \right], 
\end{align}
where $z = m_{\psi_1}/T$ and $Y_I$ ($Y_L$) is the yield for $\psi_I^{}$ (lepton number). 
The values of $Y_I^{\rm eq}$ and $Y_{\rm rel}^{\rm eq}$ respectively denote the yields for $\psi_I$ and a relativistic SM lepton given in the thermal equilibrium:
\begin{align}
Y_I^{\rm eq} = \frac{45}{2\pi^4}\frac{z^2}{g_*} {\cal K}_2\left(z\,\frac{m_{\psi_1}}{m_{\psi_I}}\right),\quad Y_{\rm rel}^{\rm eq} = \frac{45}{2\pi^4}\frac{3}{2g_{*}}\zeta(3),
\end{align}
where $\zeta(3)\simeq 1.2$ is the zeta function.  

In order to show the typical behavior of $\epsilon_I$ and $Y_B$, we consider the following simplified input parameters
\begin{align}
&(M_1,M_2,v_\varphi,m_\eta,m_\chi,m_\varphi) = (10,15,1,0.3,0.3,0.3)~\text{TeV}, \notag\\
&y_L^{11} = y_R^{11} = e^{-i\pi/4},~y_L^{12} = -y_R^{12} =: y_{12},  \label{input} \\
&y_L^{22} = y_R^{22} =: y_{22},~y_\eta^{i1} = r y_\eta,~y_\eta^{i2} = y_\eta. \notag
\end{align}
As mentioned above, the magnitude of $y_\eta$ should be of order $10^{-4}$--$10^{-5}$ to reproduce the active neutrino masses and to avoid too strong washout. 
For $y_{22} \ll 1$, we obtain $m_{\psi_3}\simeq m_{\psi_4} \simeq M_2$ with $(m_{\psi_4} - m_{\psi_3})/M_2 \ll 1$. 
In this case, $\epsilon_{3,4}$ are significantly enhanced from the second diagram in Fig.~\ref{fig:diagram2} with $\psi_{I,J} = \psi_{3,4}$ and $\psi_K = \psi_{1,2}$
due to the resonance between $\psi_3$ and $\psi_4$ as well as 
the large CPV effect coming from the larger Yukawa coupling $y_{L,R}^{11}$. 
On the other hand, the $\varphi$-loop contribution to $\epsilon_{1,2}$ are kinematically suppressed, so that the $\eta$-loop contribution is dominated. 
Therefore, we obtain $|\epsilon_{3,4}| \gg |\epsilon_{1,2}|$.

In Fig.~\ref{fig:num1}, we show the 
contour plot for the value of $|\epsilon_3|$ as a function of 
$y_{22}$ and $y_{12}$. We note that $\epsilon_{1,2}\sim 0$
and $\epsilon_4 \sim \epsilon_3$. 
As expected, larger $|\epsilon_3|$ is realized for smaller $y_{22}$, 
because the resonant effect of $\psi_3$-$\psi_4$ becomes stronger. 
This enhancement is, however, terminated at some values of $y_{22}$ depending on 
the value of $y_{12}$, because of the effect of the finite width of $\psi_{3,4}$.
We also see that the dependence on $y_{12}$ is also significant, which 
determines the size of the connection between the $\psi_{1,2}$ sector and the 
$\psi_{3,4}$ sector. 
We find that the $\epsilon_{3,4}$ parameters can be of order 1 at, for instance, $(y_{12},y_{22})\sim (10^{-1.5},10^{-5})$. 

In Fig.~\ref{fig:num2}, we show the contour plot for $|Y_B|/Y_B^{\rm obs}$ as a function of 
$r$ and $y_\eta$, where $Y_B^{\rm obs} = 8.7\times 10^{-11}$~\cite{ParticleDataGroup:2022pth} is the observed value of the present baryon number of the Universe. 
For a fixed value of $r$, we see that $|Y_B|$ significantly becomes larger for smaller $y_\eta$, because the sizable out-of-equilibrium decay of $\psi_{3,4}$ is realized. 
We also see that smaller $r$ gives larger $|Y_B|$. This is because for $r > 1$ the decays of $\psi_{1,2}$ can be more active than those of $\psi_{3,4}$ , i.e., $K_{1,2} > K_{3,4}$, so that 
the produced lepton number from the former decay is washed out by the latter with negligibly small $\epsilon_{1,2}$. 
For $r \ll 1$, such washout does not happen as the decays of $\psi_{1,2}$ are already decoupled from the thermal equilibrium, and thus the produced lepton number from 
the decays of $\psi_{3,4}$ is kept. 


Finally, we would like to show a concrete benchmark point which satisfies the observed neutrino oscillation data and the baryon asymmetry as follows: 
\begin{align}
y_\eta &= 
\begin{bmatrix}
(3.89 + 1.84 \,i) 10^{-6}  & (-7.38 + 1.10 \,i) 10^{-4} \\
(-8.58 + 4.79 \,i) 10^{-5} & (-3.58 + 1.44 \,i) 10^{-5}\\
(1.63 + 3.02 \,i) 10^{-5}  & (5.43 - 12.8 \,i) 10^{-4}
\end{bmatrix},\notag\\
\mu_1 &= 19.3~\text{GeV},~~\mu_2 = 22.1~\text{GeV}, 
\end{align}
while all the other inputs are taken to be the same way as in Eq.~(\ref{input}). 
We then obtain $Y_B = 8.6 \times 10^{-11}$ and 
\begin{align}
& \Delta m_{21}^2 = 6.94\times 10^{-5}~\text{eV}^2,~~
  \Delta m_{31}^2 = 2.51\times10^{-3}~\text{eV}^2, \notag\\
& s_{\{12, 23, 13 \}} =\{0.524,~ 0.783,~ 0.143\},~
\delta_{\rm CP}  = 117^\circ , \label{output}
\end{align}
where $s_{ij}$ indicates $\sin \theta_{ij}$, and all the values given in Eq.~(\ref{output}) are within the 3$\sigma$ range of the global fit results~\cite{Esteban:2020cvm}. 
We check that the prediction of the lepton flavor violating decays given in the above benchmark is much smaller than the current upper limit. 
For instance, the branching ratio of the $\mu\to e\gamma$ decay is given to be $\mathcal{O}(10^{-30})$ due to the small $y_\eta$ values. We also find $Z_2$ odd neutral scalar masses and mixing as 
$\{m_{H_1}, m_{H_2}, m_{A_1}, m_{A_2}\} =  \{326, 299, 301, 271\}$ GeV and $\sin \theta_{A} = -\sin \theta_{H} \simeq 0.1$.

{\it Discussions and Conclusions --} We briefly discuss dark matter physics in the model. The dark matter candidate in our model is the lightest $Z_2$ odd scalar boson since new fermions $\psi_I^{}$ are heavier to 
realize the successful leptogenesis scenario discussed above. 
For example, in our benchmark, the lightest one is $A_2$ that dominantly comes from the imaginary component of $\chi$. 
The scalar boson $A_2$ interacts with the SM gauge bosons similar to scalar dark matter given in the scotogenic model, but the coupling is suppressed by the factor $\sin \theta_A$. 
The annihilation cross section via electroweak processes, $A_2 A_2 \to W^+W^-/ZZ$, is typically 
given by $\langle \sigma v \rangle \sim 10^{-2}\times (\sin \theta_A/0.1)^4\times  (100 \ {\rm GeV}/m_{A_2})^2 \ {\rm pb}$~\cite{Barbieri:2006dq}. 
Thus, the cross section is too small for $\sin \theta_A = 0.1$ to explain the observed dark matter relic density, i.e., $\Omega h^2 \sim 0.12$~\cite{Planck:2018vyg}, 
which corresponds to $\langle \sigma v \rangle \sim 0.1$ pb~\cite{Kolb:1990vq}. 
We can, however, accommodate the observed relic density from the annihilation process via $A_2 A_2 \to Z'Z'$ with $Z'$ being the $U(1)_X$ gauge boson. 
The annihilation cross section is roughly given by $\langle \sigma v \rangle \sim 3 (g_X\cos\theta_A)^4/(64\pi m^2_{A_2})
\sim 0.1\times (g_X^{}\cos\theta_A/0.2)^4 \times  (300 \ {\rm GeV}/m_{A_2})^2  \ {\rm pb}$ with $g_X^{}$ being the $U(1)_X$ gauge coupling, so that 
$\Omega h^2 \sim 0.12$ can be reproduced by taking $g_X \simeq 0.2$ with $\cos\theta_A \simeq 1$.
Regarding the constraint from dark matter direct detections, gauge interactions only induce inelastic scatterings between dark matter and nucleus at tree level, whose cross section is negligibly small in our benchmark point with 
a mass difference of ${\cal O}(10)$ GeV between dark matter and the other $Z_2$-odd scalars. 
Although the process via the Higgs portal interaction can be important, such a coupling can be taken to be appropriately small to avoid the current upper limit on the cross section. 

Finally, let us mention the collider phenomenology to test our scenario. 
One of the promising signatures would be $pp \to \varphi \to Z'Z' \to 4\ell$ at LHC with $\ell$ being $e^\pm$, $\mu^\pm$, 
which can be realized when the mass of $\varphi$ is larger than twice the $Z'$ mass. 
The $Z'$ boson can decay into a pair of SM fermions via the kinetic mixing term in the Lagrangian, and the branching ratio of $Z' \to \ell^+\ell^-$ can be sizable if 
the $Z'$ mass is a few 100 MeV. 
We leave more detailed phenomenological studies including dark matter physics and collider analyses as future projects~\cite{MNY}. 

In conclusion, we have proposed a simple model at TeV scale which can explain neutrino oscillations, stability of dark matter and baryon asymmetry of the Universe via leptogenesis
from the common origin: the spontaneous breaking of the $U(1)_X$ symmetry. 
We have shown in Figs.~\ref{fig:num1} and \ref{fig:num2} the typical orders of the magnitude for Yukawa couplings that are required for the successful leptogenesis scenario and 
generation of neutrino masses, and then we have presented a concrete benchmark point satisfying the neutrino oscillation data, dark matter data and the observed baryon asymmetry of the Universe. 

 \vspace*{4mm}

{\it Acknowledgments --} 
We would like to thank Prof. Tetsuo Shindou for fruitful discussions about leptogenesis. 
The work was supported in part 
by National Research Foundation of Korea (NRF) Grant No. NRF-2019R1A2C3005009 (T.~M.), 
by the Fundamental Research Funds for the Central Universities (T.~N.), 
and also by the Grant-in-Aid for Early-Career Scientists, No.~19K14714 (K.~Y.).

\bibliography{references}

\end{document}